\documentclass[10pt]{iopart}
\usepackage{epsfig}
\usepackage{cite}

\def\ariadne{\textsc{Ariadne}}
\def\lepto{\textsc{Lepto}}
\def\pythia{\textsc{Pythia}}
\def\rapgap{\textsc{Rapgap}}

\begin{document} 
\jl{4}

\title[Rapidity gaps from soft colour exchanges]{Rapidity gaps at
HERA and the Tevatron from
soft colour exchanges\footnote[1]{Presented by R.\ Enberg at
UK Phenomenology Workshop on Collider Physics, Durham, England,
September 1999}}

\author{Rikard Enberg, Gunnar Ingelman and Nicusor Timneanu}

\address{High Energy Physics, Uppsala University\\
Box 535, S-751 21 Uppsala, Sweden, enberg@tsl.uu.se}

\begin{abstract} 
Models based on soft colour exchanges to rearrange colour strings in
the final state provide a general framework for both diffractive and
non-diffractive events in $ep$ and hadron-hadron collisions. We study
two such models and find that they can reproduce rapidity gap data from
both HERA and the Tevatron. We also discuss the influence of parton
cascades and multiple interactions on the results.
\end{abstract}

\noindent The rapidity gap events observed at HERA have traditionally been
explained by Regge-inspired models based on hard scattering of a parton in 
a pomeron, emitted from one of the protons \cite{IS}. These models work
well in reproducing HERA data, but in recent years rapidity gap events
have also been observed at the Tevatron 
\cite{CDF-W,CDF-JJ,Goulianos,CDF-DP,D0-DP,GoulianosWhitmore,CDF-B}. 
These data cannot be reproduced using a pomeron model with
parametrizations of the pomeron structure function from HERA data 
\cite{GoulianosWhitmore,Alvero}. This may indicate that the pomeron
approach is not universal, e.g. that the pomeron flux factorization
is not so simple. It is also clear that perturbative 
QCD cannot fully describe the formation of rapidity gaps, since they 
involve the soft part of the event, with a long space-time scale.

A different approach has been developed \cite{SCI,GAL}, where the 
non-perturbative dynamics is modelled with soft colour exchanges, 
leading to variations in the topology of the 
confining colour field, e.g.\ described by Lund strings \cite{Lund}. 
There are two models, which are similar in spirit but 
different in details: the Soft Colour Interaction (SCI) model \cite{SCI} 
and the Generalized Area Law (GAL) model \cite{GAL}. In the former, 
the partons emerging from the hard scattering are assumed to interact 
softly with the colour field of the proton, whereas in the latter, 
there are soft colour exchanges between overlapping strings. 
In both cases, the colour interactions lead to rearrangements of 
the colour charges and thereby the string topology. 
Given the softness of these interactions, changes in momenta are neglected.

The rearrangements, or reconnections, of the strings may lead to phenomena 
such as rapidity gaps, leading protons or leading neutrons. Another effect 
is that a $c\bar c$ pair may turn into a colour singlet and form a 
charmonium state \cite{SCI-charm}.

\begin{figure}[tb]
\begin{center}

\begin{tabular}{ccc} 
\begin{tabular}{l} 
\epsfig{width=0.25 \columnwidth,file=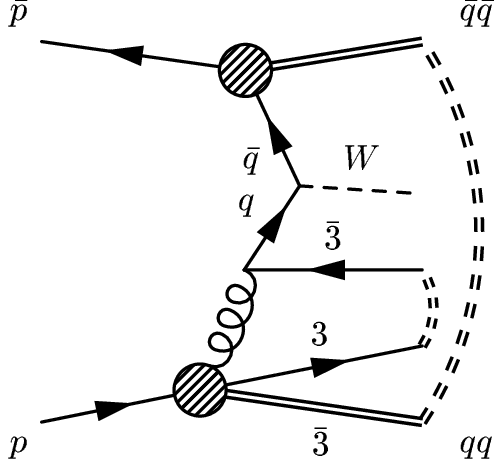,
bbllx=13, bblly=658, bburx=158, bbury=794} \\ 
\epsfig{width=0.25 \columnwidth,file=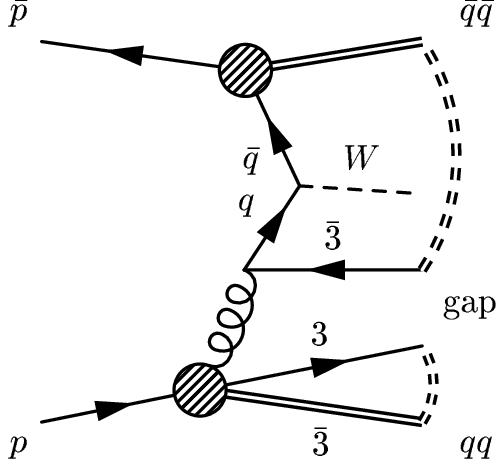,
bbllx=13, bblly=658, bburx=158, bbury=794} \\ 
\end{tabular}
&
\epsfig{width=0.3 \columnwidth , file=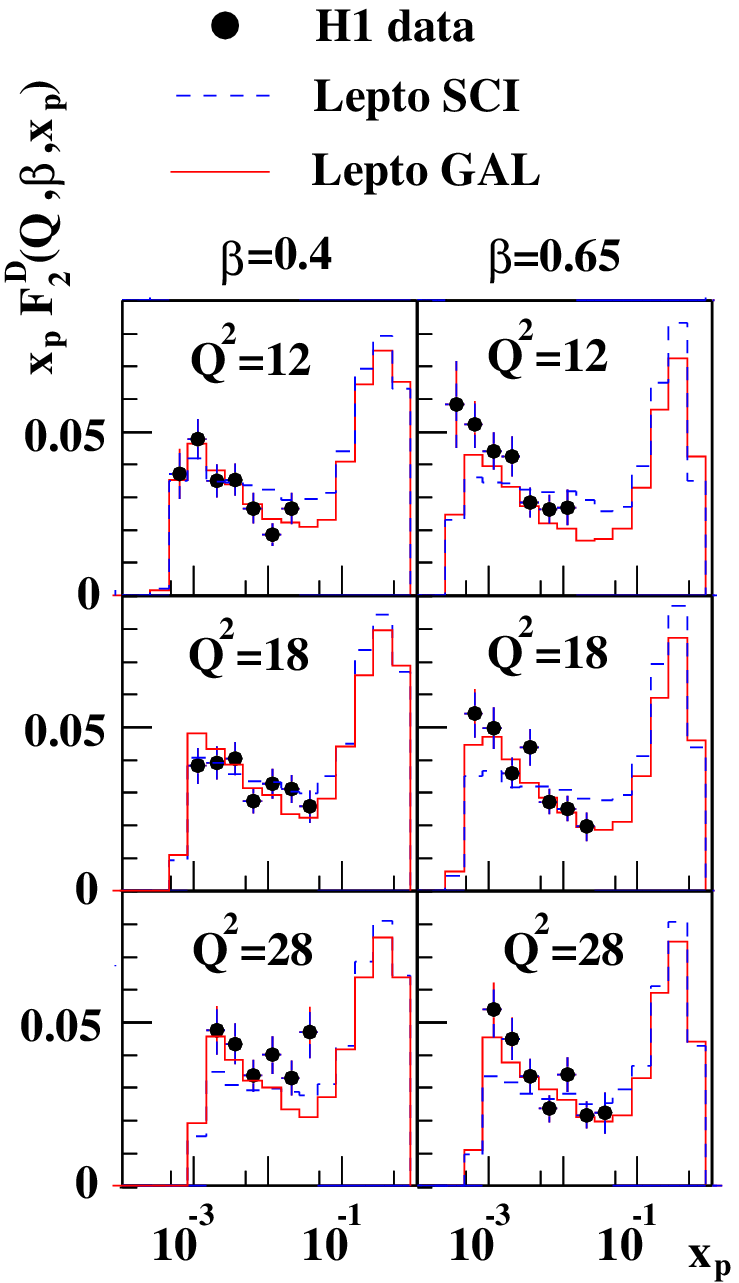,
bbllx=0, bblly=160, bburx=212, bbury=369} 
&
\epsfig{width=0.3 \columnwidth , file=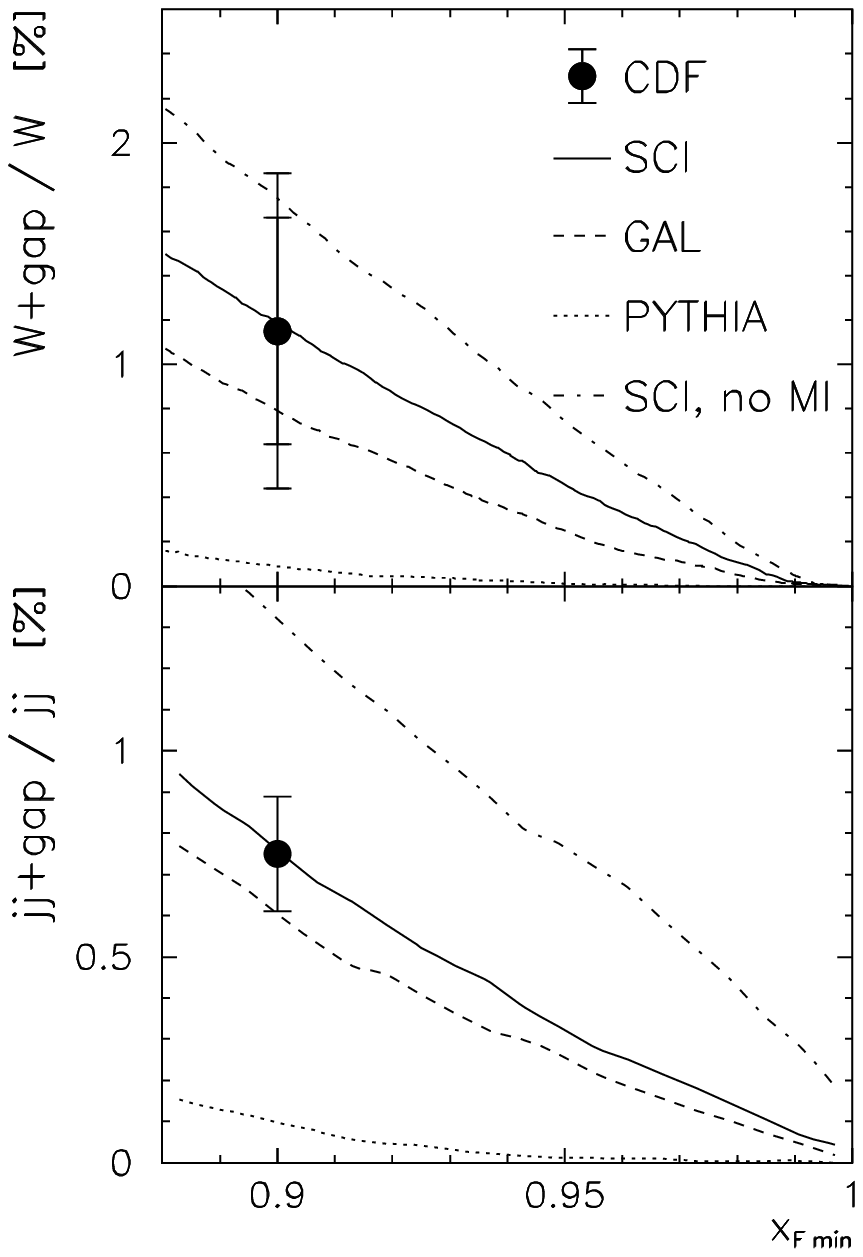,
bbllx=0, bblly=194, bburx=256, bbury=369} 
\\ (a) & (b) &(c)
\end{tabular}

\end{center}
\caption{(a) $W$ production in $p\bar{p}$ with string topology before
and after colour rearrangement. Comparison of the SCI and GAL models
to (b) the diffractive structure function $F_2^{D(3)}$ as measured
by H1 \protect\cite{H1F2D3} (plot from \protect\cite{SCI-HERA}),
and (c) the ratio of diffractive to total cross-section for $W$ 
production \protect\cite{CDF-W} and di-jet production
\protect\cite{CDF-JJ}
as measured by CDF.} \label{fig:Wdiagram}\label{fig:F2D3}
\end{figure}

The initial colour order of the partons in the final state is given by the
planar approximation in perturbative QCD, but this colour order may now be
changed by the soft colour exchanges. In \fref{fig:Wdiagram}a, we show how
this can give rise to diffractive $W$ production in $p \bar p$ collisions
at the Tevatron. The upper diagram shows the standard QCD string
configuration, with strings spanning the entire rapidity region. The lower
diagram shows the situation after a reconnection of the strings,
where there is a region in rapidity not covered by a string, such that a 
rapidity gap arises after hadronization. It is an important feature of the 
models that there is no sharp distinction between diffractive and 
non-diffractive events, but a smooth transition between the two types 
of events.

The two models have been implemented in the Monte Carlo event
generators \lepto\ \cite{LEPTO} for DIS and \pythia\ \cite{PYTHIA}
for $p\bar p$-collisions. This makes it possible to take an experimental
approach and classify events as diffractive or non-diffractive depending 
on the characteristics of the final state. The models are introduced in 
the programs as a loop over all pairs of partons (SCI) or strings (GAL) 
and each pair is allowed to exchange colour with a certain probability 
given by the free parameter $R$, which cannot be calculated in 
perturbation theory. In the SCI model, $R$ is taken as a constant and 
fitted to HERA data \cite{SCI}. In the GAL model, 
$R=R_0 [1-\exp (-b \Delta A)]$ depends on the change $\Delta A$ that 
the string rearrangement introduces in the generalized `area' swept 
by the strings in energy-momentum space \cite{GAL}. $R_0$ is a free 
parameter, simultaneously fitted to HERA and LEP data.
 
Both the SCI model and the GAL model give good descriptions of 
diffractive HERA data, e.g.\ the diffractive structure function 
in \fref{fig:F2D3}b, but also of more detailed properties of the
diffractive $X$-system \cite{SCI-HERA,SCI,GAL}. However, the description 
of inclusive final states is not as good for the SCI model, which gives 
too many soft particles, whereas the GAL model clearly improves this 
situation \cite{SCI-HERA}. The reason is that the GAL model suppresses 
strings associated with a large area (`long strings') while the SCI 
models allows reconnections giving long zig-zag-shaped strings.

Diffraction in $p\bar p$ collisions at the Tevatron offers
a new testing ground for these models. We have applied both the SCI 
and the GAL models to $p\bar p$ collisions \cite{inpreparation} in 
order to investigate $W$ and di-jet production in association with a 
rapidity gap, and also central di-jet production with two rapidity gaps.

Both models reproduce well the rates observed at the Tevatron for 
production of diffractive $W$ and diffractive di-jets when the $R$-value 
obtained from HERA data is used (see \fref{fig:F2D3}c and table
\ref{table}). The double gap fraction, i.e.\ the fraction of events 
with two gaps, conventionally associated with double pomeron exchange, 
also comes out in decent agreement with data. This should be contrasted 
to the pomeron model, which, when tuned to 
HERA data, overestimates the rates of diffractive di-jets by a factor 
6 for single gap events and 275 for double 
gap events \cite{GoulianosWhitmore}.

\begin{table}
\caption{Experimental rates (diffractive/inclusive)
measured at the Tevatron and the results obtained
from the SCI and GAL models for these rates.} \label{table}
\begin{indented}
\lineup
\item[]\begin{tabular}{@{}lllllll} \br
Observable & 
Expt & $\sqrt{s}$ & Ref. & 
Rate (\%) &
SCI  & 
GAL  \\ 
\mr

$W$  - gap & CDF   &1800 &\cite{CDF-W}    & $1.15 \pm 0.55$ & 1.2 & 0.8 \\
$jj$ - gap & CDF   &1800 &\cite{CDF-JJ}   & $0.75 \pm 0.10$ & 0.7 & 0.6 \\ 
$jj$ - gap & D\O\0 &1800 &\cite{Goulianos}& $0.76 \pm 0.08$ & 0.8 & 0.7 \\ 
$jj$ - gap & D\O\0 &630  &\cite{Goulianos}& $1.11 \pm 0.23$ & 0.9 & 1.2 \\ 
gap - $jj$ -gap$^\mathrm{a}$&CDF&1800&\cite{CDF-DP}&
$0.26 \pm 0.06$&0.2&---$^\mathrm{b}$\\
gap - $jj$ -gap$^\mathrm{a}$& D\O\0&1800&\cite{D0-DP}&(not published)
&0.2& 0.1 \\

\br
\end{tabular} 
\item[]$^\mathrm{a}$ Ratio of 2-gap events to 1-gap events. 
\item[]$^\mathrm{b}$ We have not yet obtained a rate for the GAL model. 
\end{indented}
\end{table}

In hadron-hadron collisions it is also important to consider the
underlying event, since additional soft activity may spoil the rapidity
gaps. We therefore used the model for multiple 
interactions \cite{Multiple} present in \pythia , where the 
underlying event activity is described by additional parton-parton 
scatterings with a minimum transverse momentum $p_\perp^{\mathrm{min}}$. 
However, the SCI model also contributes to the soft underlying event. 
To avoid `double counting', we decrease the amount of multiple 
interactions by increasing the free parameter $p_\perp^{\mathrm{min}}$ by 
about 500 MeV to 2.5 GeV, so that the net activity in the event will be 
the same. Having no multiple interactions (no MI in \fref{fig:Wdiagram}c)
gives a higher gap rate due to an unrealistically low event activity. 
Because of the suppression of `long' strings, the GAL model contributes 
less to the underlying event and $p_\perp^{\mathrm{min}} \simeq 2.0$ 
GeV can be used. \pythia\ used with the SCI/GAL models results in 
essentially the same jet profiles, rapidity plateaux, and particle 
multiplicities as default \pythia . 
 
Another phenomenon that has been observed is central rapidity gaps 
between jets, so that there is a hard momentum transfer across the 
gap. We find \cite{inpreparation} that the gap rates obtained using 
the GAL model as above are in agreement with the Tevatron data, but 
there are uncertainties related to multiple interactions. The SCI 
model does give such events too, but the rate is too low.

 
A potential problem is that 
the rate of gap events also depends on the amount of perturbative
emissions. It is known that the DGLAP evolution scheme does not give 
enough perturbative gluon emission in the forward region of small-$x$ 
DIS events at HERA \cite{H1forward}. This may make SCI and DGLAP
in \lepto\ overestimate the gap rate. Using SCI or similar 
models \cite{LLGap} in \ariadne\ \cite{ariadne}, which gives a 
better description of hard emissions in the forward region, gives a 
too low gap rate unless the cut-off parameter
in the cascade is increased. Similarly, the $R$ parameter in SCI and GAL
can be increased to give a larger gap rate in case of more perturbative 
emissions. 

In this context, one should realize that these cascade models are not 
very well founded theoretically. DGLAP is derived in perturbative QCD
for the inclusive case, whereas the cascade in \ariadne\ is given by a 
dipole approximation not based on Feynman diagrams. Although one may 
expect that they give fair descriptions of the mean behaviour, there
is no guarantee that they account for the fluctuations in the 
perturbative QCD emissions. Downwards fluctuations in the number of gluons 
are important for rapidity gap formation. It is therefore premature 
to draw too strong conclusions about this problem, which needs further 
investigations. To this end, we have during this workshop started to 
implement the SCI and GAL models in the \rapgap\ Monte Carlo \cite{rapgap}
which uses a model for resolved photons to describe forward parton
emissions.

In conclusion, the SCI and GAL models can give satisfactory descriptions 
of rapidity gap events in both $ep$ and $p\bar{p}$ collisions. 
They also reproduce many features of non-gap events, such that 
a unified description of both diffractive and non-diffractive 
interactions are obtained. Although these models are simple and leave some 
problems unsolved, they may represent a new way to improve our 
understanding of non-perturbative QCD dynamics.

\ack
We thank Anders Edin, Leif L\"onnblad and Johan Rathsman for helpful 
discussions.

\section*{References}

\end{document}